\documentclass[aps,amsmath,superscriptaddress,showpacs,twocolumn,prl]{revtex4}

\usepackage{graphicx} 
\usepackage[caption=false]{subfig}
\usepackage{mathptmx} 

\newcommand{\fig}[1]{\label{fig:#1}}
\newcommand{\reffig}[1]{Fig.~\ref{fig:#1}}
\newcommand{\reffigure}[1]{Figure~\ref{fig:#1}}

\newcommand \eps{\varepsilon}
\newcommand \half{\frac{1}{2}}

\begin{document}

\title{Pairing Dynamics of Polar States in a Quenched $p$-wave Superfluid Fermi Gas}

\author{Sukjin Yoon}
\email{sjyoon@ibs.re.kr}
\affiliation{Center for Theoretical Physics of Complex Systems, Institute for Basic Science, Daejeon 34051, Korea}
\affiliation{Asia Pacific Center for Theoretical Physics, Pohang, Gyeongsangbuk-do 37637, Korea}
\affiliation{Quantum Universe Center, Korea Institute for Advanced Study, Seoul 02455, Korea}

\author{Gentaro Watanabe}
\email{gentaro@zju.edu.cn}
\affiliation{Department of Physics and Zhejiang Institute of Modern Physics, Zhejiang University, Hangzhou, Zhejiang 310027, China}
\affiliation{Center for Theoretical Physics of Complex Systems, Institute for Basic Science, Daejeon 34051, Korea}
\affiliation{University of Science and Technology, Daejeon 34113, Korea}
\affiliation{Asia Pacific Center for Theoretical Physics, Pohang, Gyeongsangbuk-do 37637, Korea}
\affiliation{Department of Physics, POSTECH, Pohang, Gyeongsangbuk-do 37673, Korea}

\date{\today}

\begin{abstract}
We study the pairing dynamics of polar states in a single species $p$-wave superfluid Fermi gas following a sudden change of the interaction strength.
The anisotropy of pair interaction together with the presence of the centrifugal barrier results in profoundly different pairing dynamics compared to the $s$-wave case. 
Depending on the direction of quenches, quench to the BCS regime results in a large oscillatory depletion of momentum occupation inside the Fermi sea or a large filling of momentum occupation.
We elucidate a crucial role of the resonant state supported by the centrifugal barrier in the pairing dynamics.

\end{abstract}

\pacs{03.75.Kk, 03.75.Ss, 05.30.Fk}

\maketitle


{\emph{Introduction.--}}
Pairing of fermions and condensation of pairs are profound phenomena that have been the subjects of intensive researches in many different areas of physics including condensed matter, nuclear matter, quark matter, and cold atomic gases \cite{review:SC,review:3He,review:nuclear,review:quark,review:fermigas}. 
While the study on the fermionic condensates has a long history, the unprecedented control over the cold gases with tunable interactions via Feshbach resonances (FRs) has mapped novel landscape of superfluidity and has provided new opportunity to investigate further the nature of these symmetry-broken states from different perspective. Although the amplitude mode, referred to as the Higgs mode \cite{review:higgs}, of the order parameter was anticipated to appear in a superconductor in response to a small perturbation in non-adiabatic regime \cite{Volkov:1973}, recent resurgent studies on the amplitude modes of order parameters have been motivated from new prospects in cold atomic gases \cite{higgs:atom} : 
the large dynamical time-scale and high controllability of the cold atomic gases have made quantum quench, sudden change of system's parameter(s) at zero temperature, become a practical subject of experimental control as well as theoretical investigation and provided unprecedented opportunities of studying the subsequent non-equilibrium dynamics.
For the last decade, the coherent dynamics of the order parameter induced by quantum quenches in $s$-wave Fermi gases has been investigated and classified \cite{Barankov:2004,Andreev:2004,Szymanska:2005,Yuzbashyan:2006,Barankov:2006,Yuzbashyan:2015,Gurarie:2009,Bulgac:2009,Scott:2012}.

Finite angular-momentum-paired atomic superfluids are compelling because they are or may be connected to superfluid $^3\text{He}$ \cite{review:3He}, neutron superfluids inside neutron stars \cite{Sauls:1989}, high-$T_c$ superconductors, and the topological features useful for the quantum computing \cite{Read:2000,Tewari:2007}. 
As with finite angular-momentum-paired superfluids, $p$-wave-paired atomic gases can exhibit distinct phases of superfluidity characterized by their rich structure of order parameters \cite{Gurarie:QPT_Pwave,Gurarie:Pwave,Cheng:2005}; these phases and the phase transitions between them can be accessible via tuning between BCS and Bose-Einstein condensation (BEC) limits with a $p$-wave FR.
In several experiments \cite{Ticknor:2004, Zhang:2004, Schunck:2005, Gaebler:2005, Fuchs:2008, Mukaiyama:2016}, $p$-wave FRs have been realized and also there have been efforts or proposals \cite{Zhang:2008,Han:2009,Cooper:2009,Williams:2012,Williams:2013,Diaz:2013,Fu:2014,Yamaguchi:2015} for achieving stable $p$-wave atomic superfluids which has been challenging due to inelastic collisional losses \cite{pwave_stability:exp,pwave_stability:theo}.
Furthermore, in $^3\text{He}$, pure polar ($p_x$) state has been recently realized below the transition temperature in uniaxially anisotropic aerogel \cite{3He:polar}. In addition, the dynamical behaviors of the order parameters of axial ($p_x + ip_y$) states in 2 dimensions has been theoretically studied in \cite{Foster:2013,Dong:2015,Dzero:2015}.

Motivated by the recent advancement, we study the pairing dynamics after a sudden change in the interaction strength for polar states of a $p$-wave superfluid Fermi gas at zero temperature. Dynamics of the pairing field as well as the momentum distribution and the pair amplitude distribution in momentum space is obtained numerically within a mean-field (MF) approach. Unlike previous studies for axial states \cite{Foster:2013,Dong:2015,Dzero:2015} in 2 dimensions, we especially consider 3 dimensional systems since the MF approach is more reliable and we can make more direct comparison between existing results on $s$-wave superfluids in 3 dimensions \cite{Barankov:2004,Andreev:2004,Szymanska:2005,Yuzbashyan:2006,Barankov:2006,Yuzbashyan:2015,Gurarie:2009,Bulgac:2009,Scott:2012}.

Our findings are as follows: 
the anisotropy of the $p$-wave paired polar states together with the presence of the centrifugal barrier gives rise to the pairing dynamics qualitatively different from the $s$-wave Fermi gas. When the interaction is quenched to BCS side either from BCS or from BEC side, a centrifugal barrier supports resonant molecular states at the final strength of the interaction. We show the crucial role of the resonant state causing the rich pairing dynamics of $p$-wave Fermi superfluids, which has not been clarified in the previous studies \cite{Foster:2013,Dong:2015,Dzero:2015}.
(i) When the interaction is changed to a value of stronger attraction within BCS regime, an oscillatory hole-burning (large depletions in momentum distribution) appears at different rate depending on the polar angle from the symmetry axis.
(ii) When the interaction is changed to a value of weaker attraction within BCS regime or changed from BEC regime to BCS regime, an oscillatory particle-peak (large filling in the momentum distribution) appears at different rate depending on the polar angle from the symmetry axis.
(iii) The decaying oscillation frequency of the pairing field agrees with hole-burning/particle-peak frequency on the symmetry axis. This is connected to the observation that pair amplitudes in the region of hole-burning/particle-peak located off the symmetry axis in momentum space display a phase structure like a vortex-ring.


{\emph {Model and Formulation.--}} 
We consider a single-species polarized Fermi gas with a $p$-wave interaction $V_1({\bf k,k'})$,
\begin{equation}
H = \sum_{\bf k} \xi_k \hat{a}_{\bf k}^{\dagger} \hat{a}_{\bf k}  +  \half \sum_{\bf k,k',q} V_{1}({\bf k,k'}) 
\hat{b}_{\bf k,q}^{\dagger} \hat{b}_{\bf k',q} \ ,
\end{equation}
where $\hat{a}_{\bf k}^{\dagger}$ creates a fermion atom of momentum ${\bf k}$, $\hat{b}^{\dagger}_{\bf k,q} = \hat{a}_{{\bf k+ q}/2}^{\dagger} \hat{a}_{{\bf -k +q}/2}^{\dagger}$, and $\xi_k = k^2/2M - \mu$ with $M$ the atom mass, $\mu$ the chemical potential, and $\hbar = 1$.
The Pauli principle excludes the possibility of $s$-wave channel interaction for a single-species Fermi gas and we introduce the dominant $p$-wave attractive interaction potential of a separable form 
\begin{equation}
V_{1} ({\bf k,k'})  = -4 \pi g \Gamma_m^{\ast}({\bf k})\Gamma_m({\bf k'})
\end{equation}
with
\begin{equation}
\Gamma_m({\bf k}) = \frac{kk_0}{k^2+k_0^2} Y_{1,m} (\hat{\bf k})  \ ,
\end{equation}
where $g$ ($>0$) is a coupling strength, $m$ is the angular momentum projection, and $k_0$ is a momentum cutoff \cite{Ho:2005,Iskin:2006}.

In the $p$-wave FR, the dipolar anisotropy splits FR into a doublet of $m=\pm1$ and $m=0$ resonances \cite{Ticknor:2004}. Due to the dipolar splitting, the system can be independently tuned into $m=0$ and $m=\pm 1$ resonances displaying axial ($p_x+ip_y$) and polar ($p_x$) states separately \cite{Gurarie:QPT_Pwave, Gurarie:Pwave, Cheng:2005, Mukaiyama:2016}. 
We consider a $p$-wave FR of $m=0$ projection:
$\Gamma_0({\bf k}) = \frac{kk_0}{k^2+k_0^2} Y_{1,0} (\hat{\bf k})$
where $Y_{1,0} (\hat{\bf k}) = \sqrt{3/4 \pi} \cos \theta$.

The low energy scattering amplitude for the $p$-wave channel 
$f_1(k) = {k^2}/ \left ( { -\frac{1}{a_p} + \frac{r_p k^2}{2} -i k^3 } \right )$
is obtained provided $g$ and $k_0$ are related to the two parameters ($a_p$ is scattering volume and $r_p$ has dimension of inverse length) :
$\frac{1}{4\pi g} = -\frac{MV}{16 \pi^2 a_p k_0^2} + \sum_{k}\frac{|\Gamma_0(k)|^2}{2 \epsilon(k)}$ and $r_p = - \frac{1}{k_0} \left( k_0^2 + \frac{4}{k_0 a_p} \right)$ \cite{Ho:2005}. The pole of the scattering amplitude is given by $E_{\text{pole}} \approx 2/Ma_p r_p$ which corresponds to a bound state for $a_p  < 0$ and a resonant state (due to a centrifugal barrier) for $a_p > 0$. The width of the resonant state is $\Gamma_p \approx E_{\text{pole}} \sqrt{32/| a_p r_p^3 |} $ \cite{Gurarie:Pwave}.

We employ the mean-field approximation for pairing between atoms of equal and opposite momenta (see \cite{Gurarie:QPT_Pwave} for an argument on the reliability of this approximation in $p$-wave Fermi gases of low density).
The time evolution of our system is described by the time-dependent Bogoliubov-de Gennes (BdG) equations
\begin{equation}
i \frac{\partial}{\partial t}
\left( \begin{array}{c} 
u_{\mathbf k} (t) \\ v_{\mathbf k} (t)
\end{array} \right)  =   
\left( \begin{array}{cc}
\xi_{\mathbf k} & \Delta_{ \mathbf k}(t) \\
\Delta_{\mathbf k}^\ast (t)& - \xi_{\mathbf k} 
\end{array} \right)
\left( \begin{array}{c} 
u_{\mathbf k} (t)\\ v_{\mathbf k} (t)
\end{array} \right)  \ ,
\end{equation}
where $u_{\bf k}(t)$ and $v_{\bf k}(t)$ are quasiparticle amplitudes and the pairing field $\Delta_{ \mathbf k}  (t)$ is given by
\begin{equation}
\Delta_{ \mathbf k}  (t)= \sum_{ \mathbf  k'} V_1({ \mathbf  k,  \mathbf k'})u_{ \mathbf  k'}^{\ast}(t) v_{ \mathbf  k'}(t)\ .
\end{equation}
In our potential of separable form,
$
\Delta_{\mathbf k} = \Gamma_0({\mathbf k}) \Delta_0 
                                 = \Gamma_0({\mathbf k}) (-4\pi g \sum_{ \mathbf  k'} \Gamma_0({\mathbf k'}) u_{ \mathbf  k'}^{\ast} v_{ \mathbf  k'}) \nonumber 
$. For convenience, we call $|\Delta_0|$ as a pairing amplitude even though the actual pairing amplitude $|\Gamma_0({\mathbf k})\Delta_0|$ has a momentum dependence. 
The anisotropic nature of $ \Gamma_0({\mathbf k}) $ plays a role in the pairing dynamics as we can see shortly.

In our discussion, we assume that gas sample size is smaller than the coherence length and neglect inhomogeneous phase fluctuations and vortices. Our initial state is prepared in the ground state at an initial scattering volume $(a_p)_{\text{init}}$ and the scattering volume is suddenly changed to a final value $(a_p)_{\text{fin}}$ in a time interval $\delta t \approx 0.01/\eps_F \ll 1/\eps_F$ where $\eps_F = k_F^2/2M$ with $k_F = (6\pi^2 N/V)^{1/3}$.


{\emph{Results.--}} 
We choose $k_0 =  40k_F$ where $k_0/k_F \gg 1$ is satisfied for the low density of the gas.
The parameter $r_p$ does not change much during the quench processes discussed below and we safely fix the value of $r_p$ at the initial value during the time evolution. 

Equilibrium properties of polar states from BCS to BEC regime were studied in \cite{Iskin:2006} and our parameters give qualitatively the same behavior;
with increasing $1/k_F^3 a_p$, we go from the BCS side ($\mu<0$) to the BEC side ($\mu>0$) separated by the unitary point ($\mu=0$) at $1/k_F^3 a_p \approx 0.5$. The asymptotic value of the pairing amplitude $|\Delta_0|$ in the BEC limit is $\Delta_{\text{BEC}} = 8 \eps_{F} (0.5 k_0^2/9\eps_{F})^{1/4} \approx 29.2\eps_F$ and $|\Delta_0|$ is almost saturated in the BEC side.


First, we consider the case that $1/k_F^3a_p$ is quenched to the value of stronger attraction (hereafter, called ``forward quench''):
(a) BCS $\to$ BCS ($1/k_F^3a_p\!: -20 \to -10$), (b) BCS $\to$ BEC ($1/k_F^3a_p \!\! : -20 \to +10$), and (c) BEC $\to$ BEC ($1/k_F^3a_p\!\!: +5 \to +20$). 
The dynamics of the pairing amplitude $|\Delta_0|$ and the momentum distribution $|v_{\bf k}|^2 (k_y =0)$ at the local maximum (in time domain) of the pairing field are shown in \reffig{polar_forward}.
The dynamics in the case of quench within BEC regime ($1/k_F^3a_p:  +5 \to +20$) is not shown because $|\Delta_0|$ is almost saturated in the BEC side and there is no noticeable pairing dynamics. 
Only the first quadrant of $k_x$\nobreakdash-$k_z$~plane is shown due to the azimuthal symmetry about $k_z$-axis and the reflection symmetry on the $k_x$\nobreakdash-$k_y$~plane.

\begin{figure}[!t]
\centering
\subfloat{
\hspace{-0.50\columnwidth}
\includegraphics[clip,width=0.46\columnwidth]{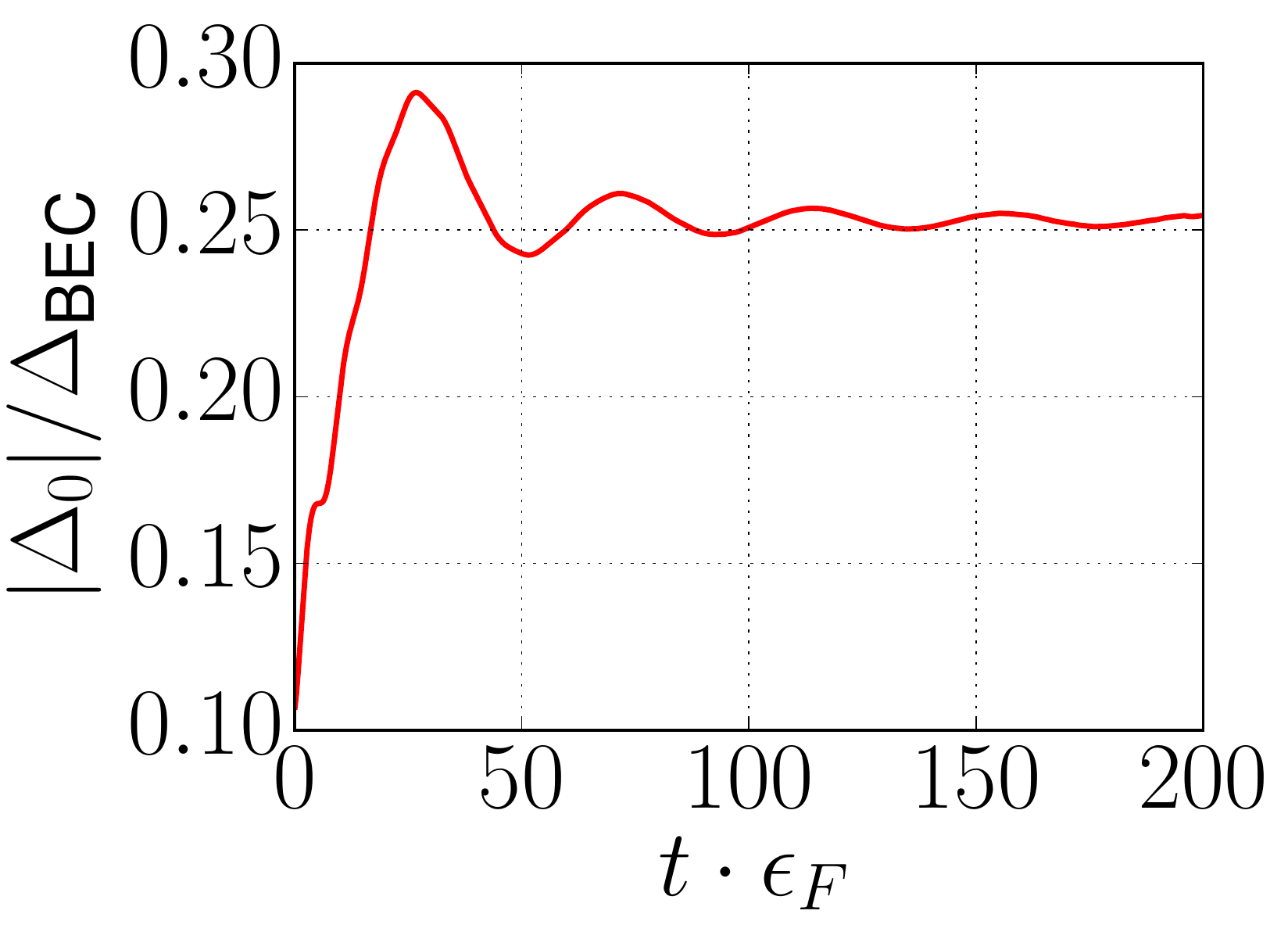}%
\hspace{0.01cm}
\includegraphics[clip,width=0.54\columnwidth]{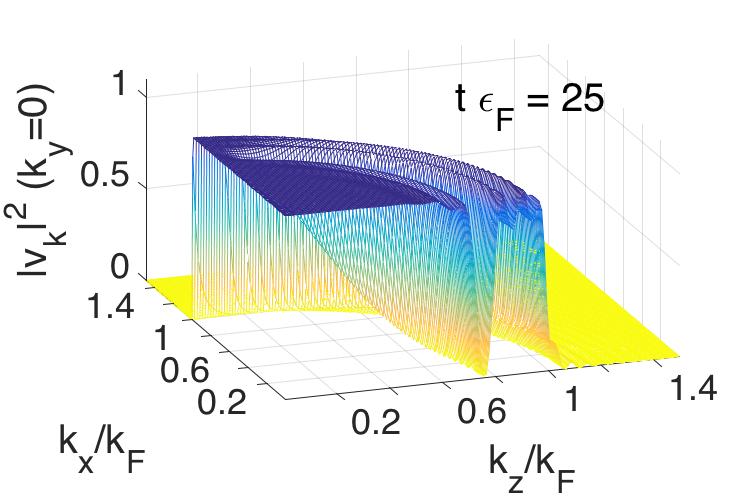}
\hspace{-0.565\columnwidth}
(a)
}

\vspace{-0.4cm}
\subfloat{
\hspace{-0.50\columnwidth}
\includegraphics[clip,width=0.46\columnwidth]{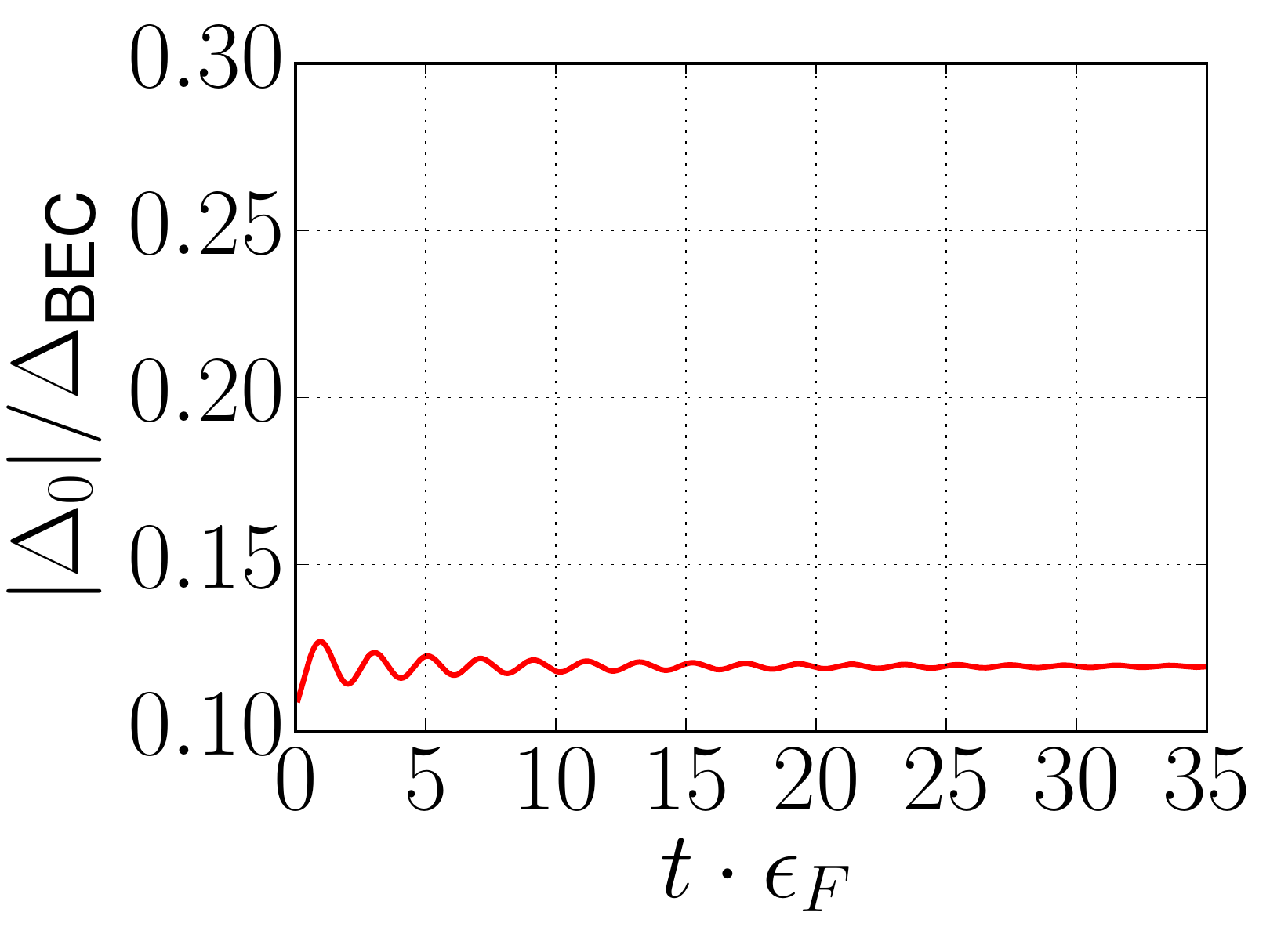}%
\hspace{0.01cm}
\includegraphics[clip,width=0.54\columnwidth]{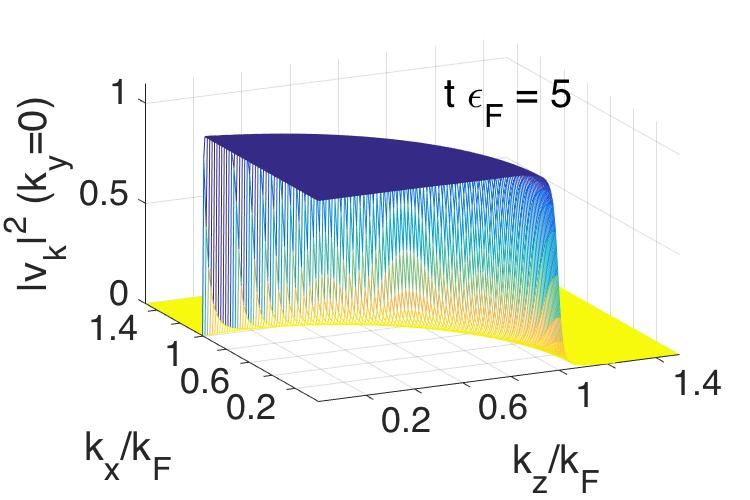}
\hspace{-0.565\columnwidth}
(b)
}
\caption{Dynamics of $|\Delta_0|$ and the momentum occupation $|v_{\bf k}|^2$ at the designated time. Inverse scattering volume $1/k_F^3a_p$ is quenched (a) from $-20$ to $-10$ and (b) from $-20$ to $10$.}
\fig{polar_forward}
\end{figure} 

In case (a), large depletion of momentum occupation inside the Fermi sea (called ``hole-burning'') is observed and it is connected to a large time-scale oscillation of $|\Delta_0|$ (on top of it, a short time-scale decaying oscillation of small amplitude is seen at the early stage up till $t = 25/\eps_F$ or so). This oscillatory hole-burning results from the interconversion between atomic and resonant states.
The hole-burning point is at $k=k_h$ satisfying $k_h^2/2M \approx E_r /2 \approx 0.5\eps_F $ where $E_r$ is the resonant state energy at the final value of $1/k_F^3 a_p$. Also, the location of the hole-burning point is independent of the initial values of the scattering volume for a given final value of the scattering volume. 

\begin{figure}[!b]
\includegraphics[clip,width=0.5\columnwidth]{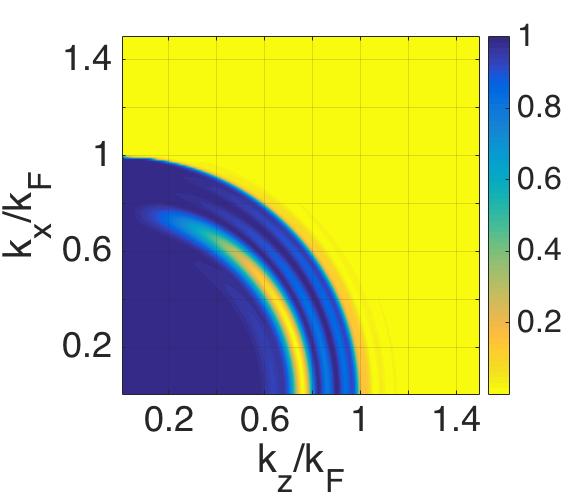}%
\includegraphics[clip,width=0.5\columnwidth]{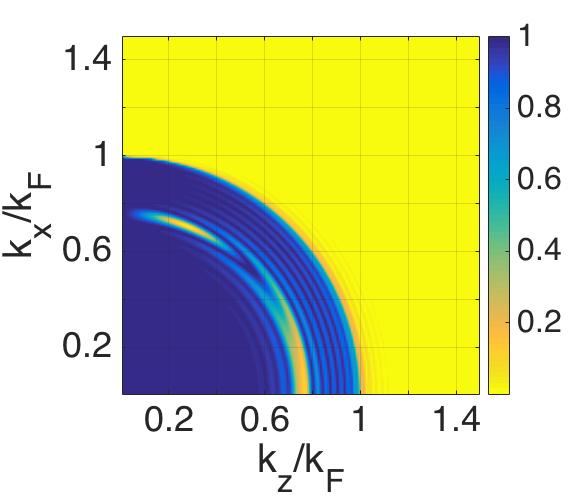}
\vspace{-0.7cm}
\caption{Momentum occupations $|v_{\bf k}|^2$ at $t \cdot \eps_F=25$ (left) and $60$ (right). $1/k_F^3a_p$ is quenched from $-20$ to $-10$.}
\fig{polar_forward_2d}
\end{figure} 

The anisotropy of the $p$-wave pairing field ($\Delta_{\bf k} (t) \propto \cos \theta $) makes the hole-burning rate slow down as the polar angle from the symmetry axis ($k_z$-axis) increases 
and hole-burning appears as in \reffig{polar_forward} (a) at the early stage of $t \cdot \eps_F=25$. 
\reffigure{polar_forward_2d} shows the density plots of case (a) in \reffig{polar_forward} at $t \cdot \eps_F=25$ (left) and $60$ (right). It shows that a hole-burning region created on the $k_z$-axis evolves into a hole-burning ring which expands in radius and moves to the equator ($k_z=0$) while another hole-burning region of smaller extent is being created on the $k_z$-axis. 

The pairing field $|\Delta_0|$ reaches a local maximum in time domain when the momentum occupation at the hole-burning point is completely depleted (i.e. hole-burning is maximum) on the $k_z$-axis and reaches a local minimum when the hole-burning on the $k_z$-axis disappears.
This agreement between the oscillation frequency of the pairing field $|\Delta_0|$ and the hole-burning frequency on the $k_z$-axis will be explained when the pair amplitude distribution is discussed.
The frequency of the hole-burning oscillation in case (a) is $\omega \approx 0.12 \eps_F$.
Compared with a $s$-wave case where the hole-burning frequency is, in the narrow resonance limit, the detuning energy which is almost molecular (resonant) state energy in the weak coupling limit \cite{Andreev:2004}, the hole-burning frequency in the $p$-wave Fermi gas is much smaller than the resonant state energy: $\omega/E_r \approx 0.12$ in case (a) above.

As much as the large time-scale oscillation of $|\Delta_0|$ is related to the oscillatory hole-burning resulting from the interconversion between atomic and resonant states, its oscillation period is qualitatively related to the width of the resonant state even though the quantitative relation is a very complicated problem beyond the discussion in this letter; the large time-scale oscillation period becomes larger as the resonance width $\Gamma$ as well as the resonant state energy becomes smaller when the final $1/k_F^3 a_p$ approaches the unitarity. 
When the final $1/k_F^3 a_p$ is put in BEC side ($\mu<0$) as in case (b), there does not appear a hole-burning: a large time-scale oscillation connected to hole-burnings disappears and there remains a decaying oscillation with a short time-scale which is connected to the small dynamics near the Fermi surface of the initial state [see \reffig{polar_forward}(b)].


\begin{figure}[!t]
\centering
\subfloat{
\hspace{-0.50\columnwidth}
\includegraphics[clip,width=0.46\columnwidth]{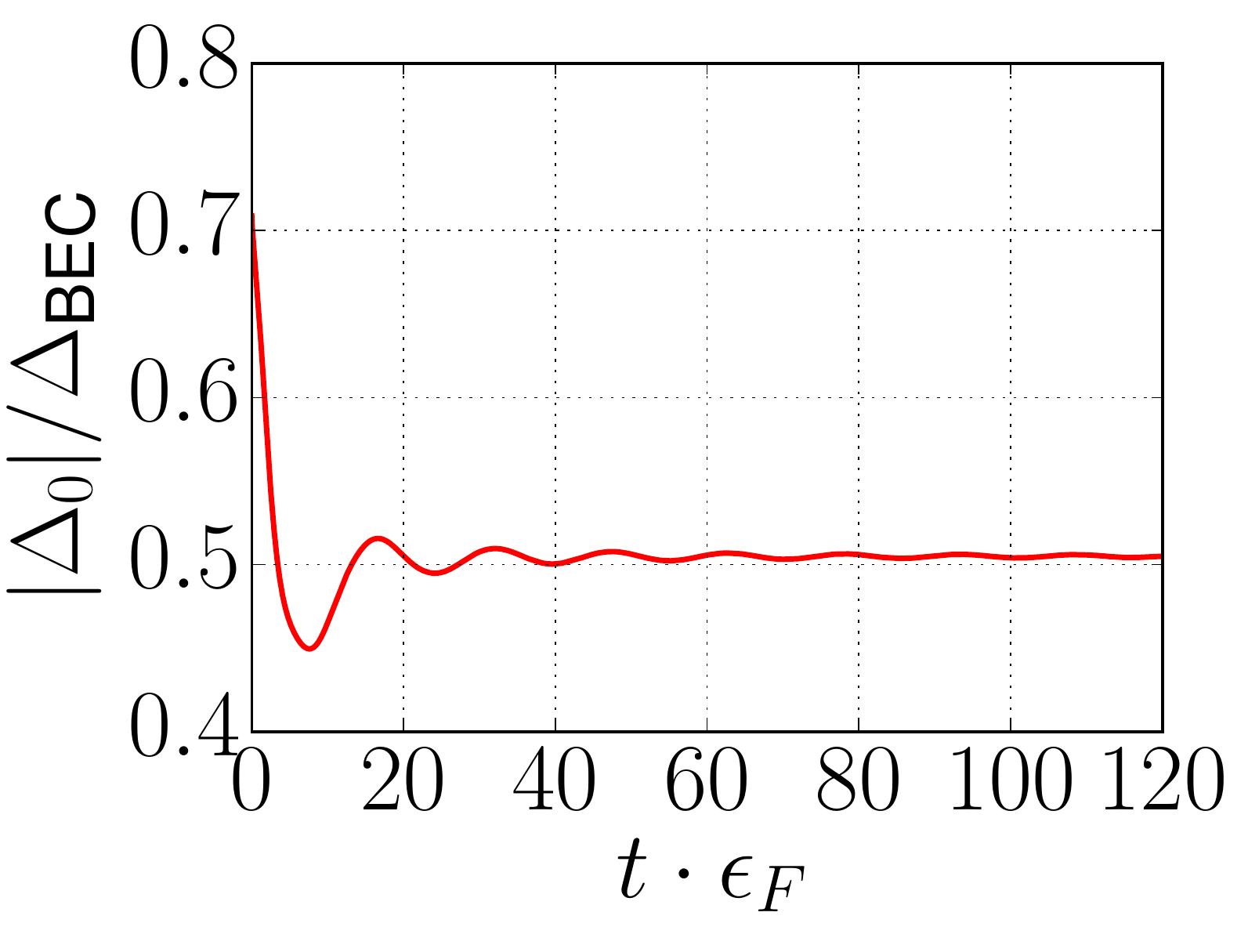}
\hspace{0.01cm}
\includegraphics[clip,width=0.54\columnwidth]{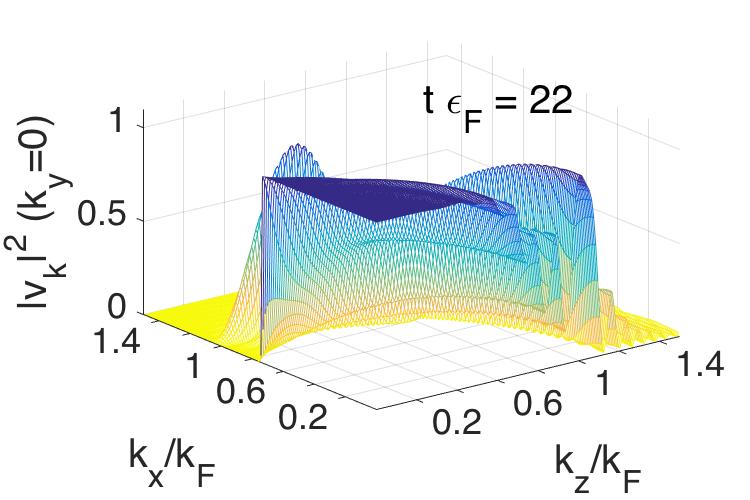}
\hspace{-0.565\columnwidth}
(a)
}

\vspace{-0.4cm}
\subfloat{
\hspace{-0.50\columnwidth}
\includegraphics[clip,width=0.46\columnwidth]{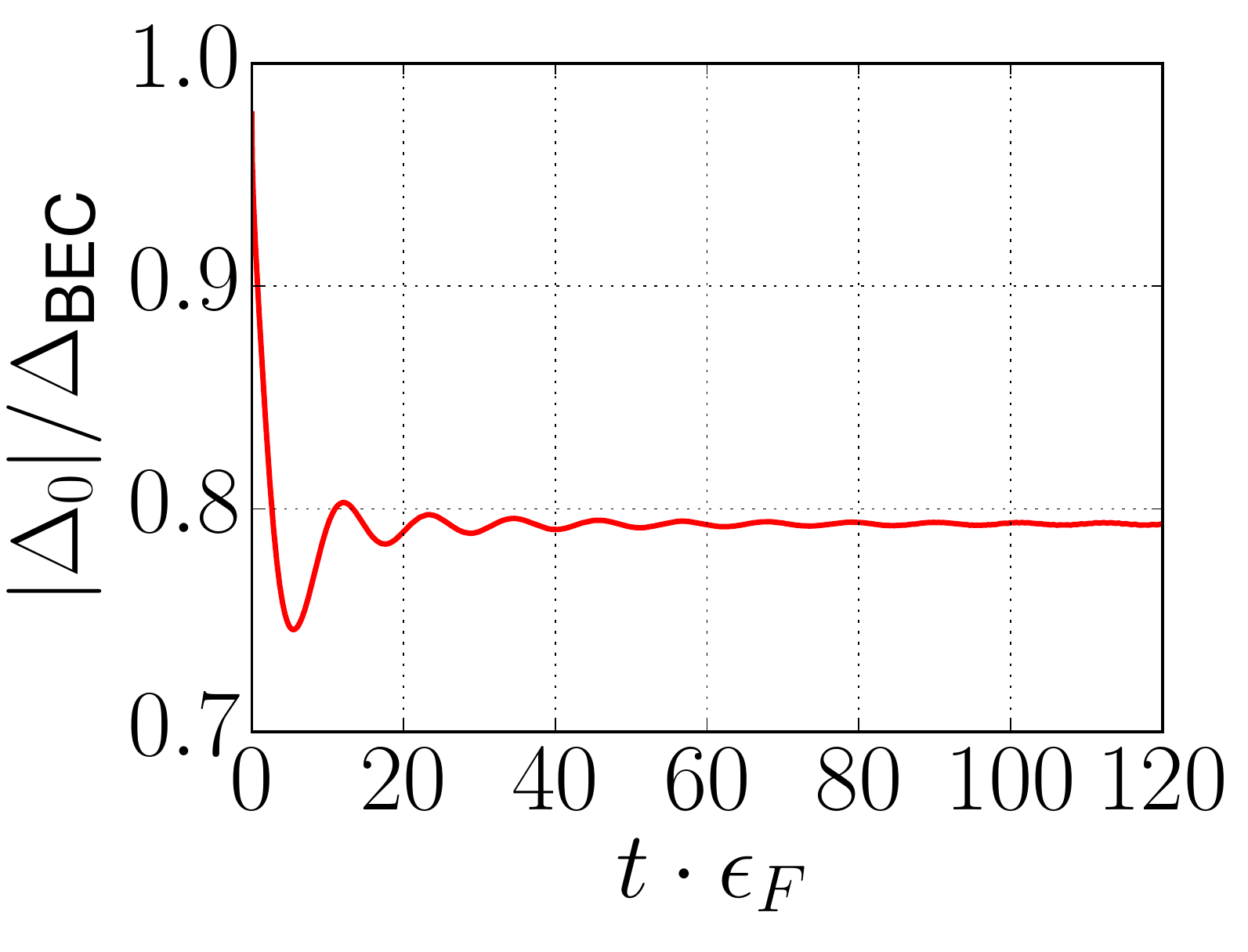}
\hspace{0.01cm}
\includegraphics[clip,width=0.54\columnwidth]{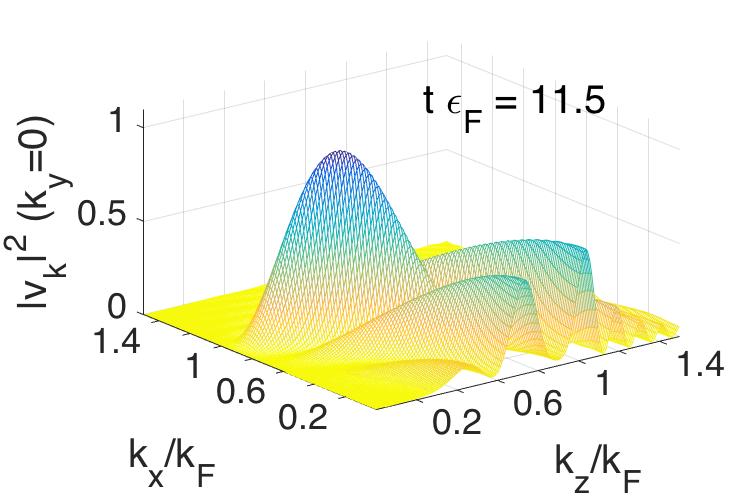}
\hspace{-0.565\columnwidth}
(b)
}
\caption{Dynamics of $|\Delta_0|$ and the momentum occupation $|v_{\bf k}|^2$ at the designated time. Inverse scattering volume $1/k_F^3a_p$ is quenched (a) from $-10$ to $-20$ and (b) from $+10$ to $-15$.}
\fig{polar_backward}
\end{figure}

Next, we consider the case that the inverse scattering volume is quenched to the value of weaker attraction (hereafter, called ``backward quench''):
(a) BCS $\to$ BCS ($1/k_F^3a_p\!: -10 \to -20$), (b) BEC $\to$ BCS ($1/k_F^3a_p\!: +10 \to -15$), and (c) BEC $\to$ BEC ($1/k_F^3a_p\!: +20 \to +5$, not shown for the same reason as in the forward quench).
\reffigure{polar_backward} shows the dynamics of the pairing amplitude $|\Delta_0|$ and the momentum distribution $|v_{\bf k}|^2 (k_y =0)$ at the designated time. The density plots of cases (a) and (b) in \reffig{polar_backward} are displayed in \reffig{polar_backward_2d}.

\begin{figure}[!t]
\hspace{-0.7cm}
\subfloat{
\raisebox{-0.8\height}{
\includegraphics[clip,width=0.5\columnwidth]{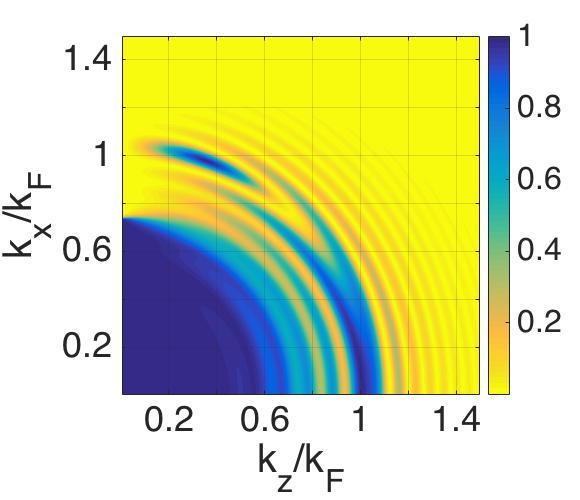}
}
\hspace{-1.5cm}(a)\hspace{0.8cm}
}
\subfloat{
\raisebox{-0.8\height}{
\includegraphics[clip,width=0.5\columnwidth]{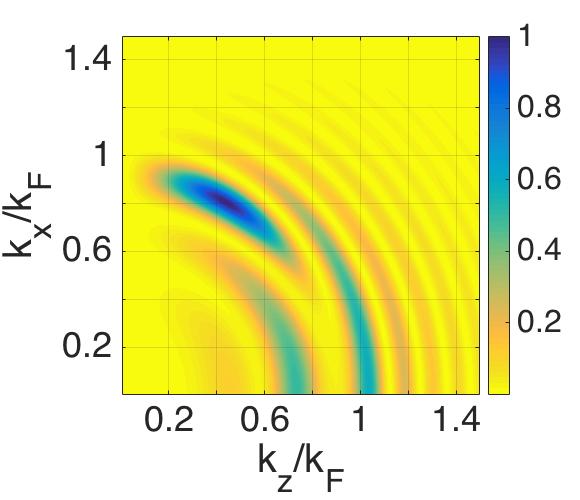}
}
\hspace{-1.5cm}(b)\hspace{0.5cm}
}
\vspace{-0.3cm}
\caption{The snapshots of the momentum occupation $|v_{\bf k}|^2$. $1/k_F^3a_p$ is quenched (a) from $-10$ to $-20$ (left) and (b) from $+10$ to $-15$ (right).}
\fig{polar_backward_2d}
\end{figure}

When quenched to BCS regime in the backward direction regardless of whether the initial state is in BCS or BEC side, there appears a peak of the quasiparticle amplitude (called ``particle-peak'') in momentum occupation $|v_{\bf k}|^2$. This is nothing but a signature of dissociation of (a) BCS-paired atoms and (b) diatomic molecules via resonant state of the final Hamiltonian.
In case (a), a particle-peak is at $k=k_p$ satisfying $k_p^2/2M \approx E_r / 2 \approx \eps_F $. The location of the particle-peak is independent of the initial states with $1/k_F^3 a_p = -15$, $-10$, and $-5$ (also for $+5$ and $+20$ which belong to (b) BEC $\to$ BCS).
In case (b), a particle-peak is also at $k=k_p$ satisfying $k_p^2/2M \approx E_r /2 \approx 0.8 \eps_F $.

The anisotropy of the $p$-wave pairing field ($\Delta_{\bf k} (t) \propto \cos \theta $) makes the dissociation rate slow down as the polar angle from the $k_z$-axis increases: a particle-peak appears on the $k_z$-axis at the early stage and moves to the equator while another particle-peak is being created on the $k_z$-axis as seen in \reffig{polar_backward} and \reffig{polar_backward_2d}. 
The pairing amplitude $|\Delta_0|$ reaches a local minimum in time domain when the particle-peak reaches a maximum on the $k_z$-axis, and it reaches a local maximum in time domain when the particle-peak disappears on the $k_z$-axis: the oscillation frequency of the pairing amplitude $|\Delta_0|$ agrees with that of the particle-peak on the $k_z$-axis.

\begin{figure}[!t]
\subfloat{
\hspace{-0.50\columnwidth}
\raisebox{-0.07\height}{
\includegraphics[clip,width=0.50\columnwidth]{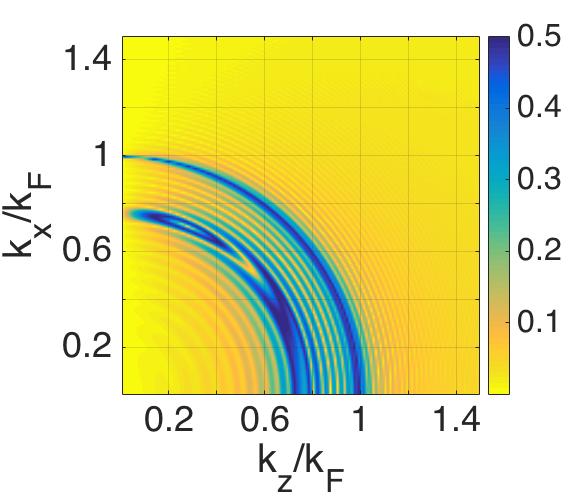}
\includegraphics[clip,width=0.50\columnwidth]{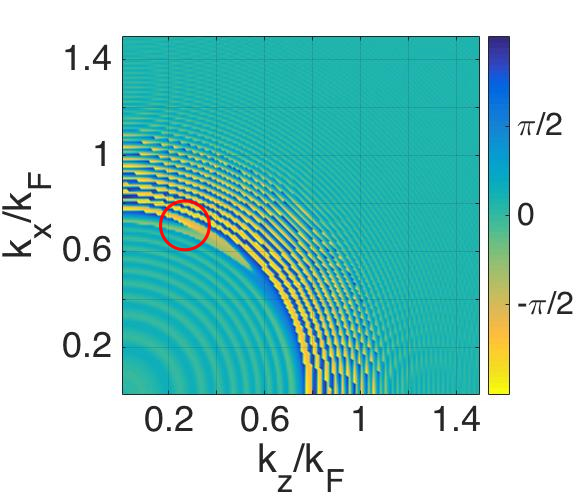}
}
\hspace{-0.55\columnwidth}
(a)
}

\vspace{-0.4cm}
\subfloat{
\hspace{-0.50\columnwidth}
\raisebox{-0.07\height}{
\includegraphics[clip,width=0.5\columnwidth]{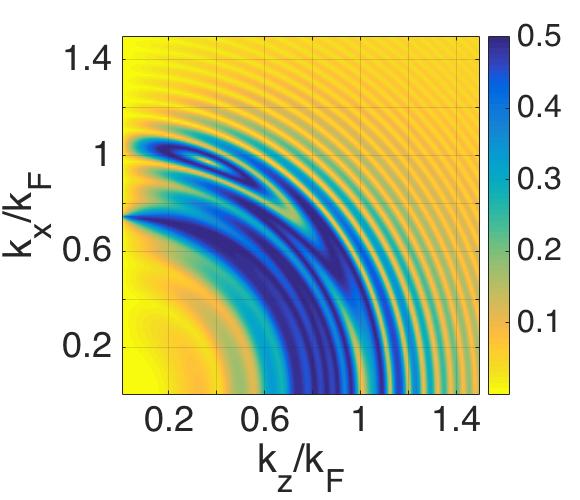}
\includegraphics[clip,width=0.5\columnwidth]{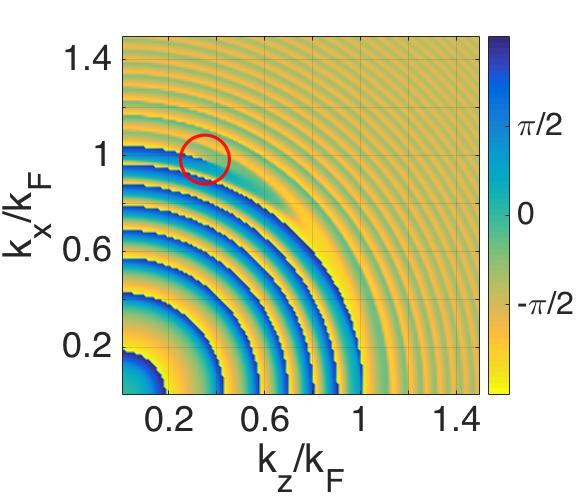}
}
\hspace{-0.55\columnwidth}
(b)
}

\vspace{-0.4cm}
\subfloat{
\hspace{-0.50\columnwidth}
\raisebox{-0.07\height}{
\includegraphics[clip,width=0.5\columnwidth]{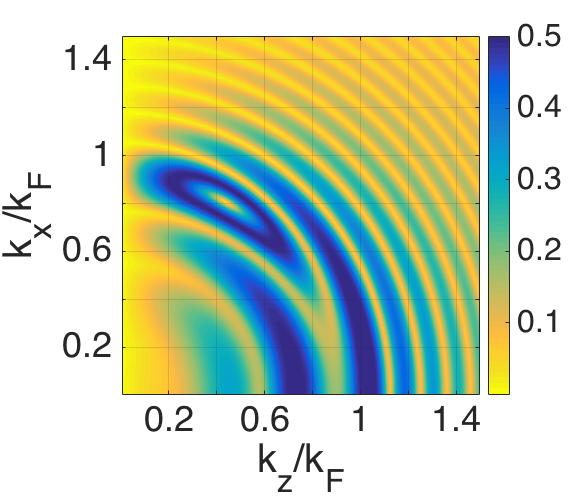}
\includegraphics[clip,width=0.5\columnwidth]{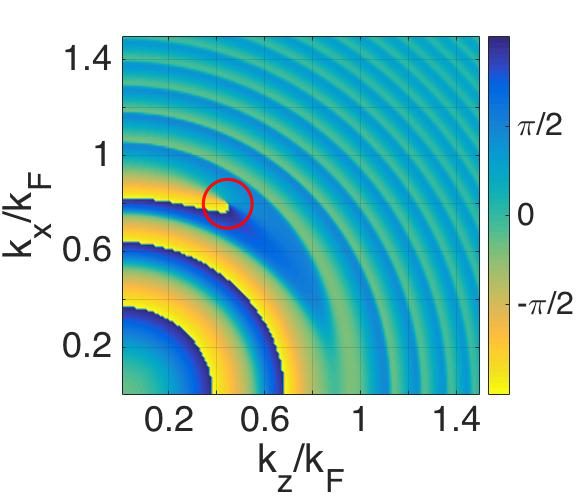}
}
\hspace{-0.55\columnwidth}
(c)
}
\caption{The snapshots of the magnitude (left column) and the phase (right column) of the pair amplitude $u^{*}_{\bf k} v_{\bf k}$ in momentum space. The red circles in right panels highlight the position of the phase rotation of $2\pi$. $1/k_F^3a_p$ is quenched (a) from $-20$ to $-10$ (top), (b) from $-10$ to $-20$ (middle), and (c) from $+10$ to $-15$ (bottom). The snapshots are taken at $t \cdot \eps_F = 60$, $22$, and $11.5$, respectively.}
\fig{polar_vortex_ring}
\end{figure} 

To understand the relation between the dynamics of the pairing amplitude $|\Delta_0|$ and the dynamics of quasiparticles in momentum space, a close look at the dynamics of the pair amplitudes $u^{*}_{\bf k} v_{\bf k}$ is helpful.  
\reffigure{polar_vortex_ring} shows the magnitude (left panel) and phase (right panel) of the pair amplitude $u^{*}_{\bf k} v_{\bf k}$ in momentum space in the cases of forward quench ($1/k_F^3a_p\!: -20 \to -10$ ) and backward quenches ($1/k_F^3a_p\!:$ $-10 \to -20$ and  $+10 \to -15$). 

The hole-burning region (ring shape around the $k_z$-axis) off the $k_z$-axis has a phase rotation of $2\pi$ around the core (see the red circles in the right panels of \reffig{polar_vortex_ring}) while the hole-burning region created on the $k_z$-axis cannot topologically. This makes the contribution of the pair amplitudes around the vortex-ring shaped region to the pairing field $\Delta_0$ become very small because
$\Delta_0 \propto  \sum_{\mathbf k'} \Gamma_0({\mathbf k'}) u_{ \mathbf  k'}^{\ast} v_{ \mathbf  k'}$ and $\Gamma_0({\mathbf k})$ takes similar values on the cross-sectional area of the hole-burning ring: there is a strong cancellation in the summation $\sum_{\mathbf k'} u_{ \mathbf  k'}^{\ast} v_{ \mathbf  k'}$ for the hole-burning region off the $k_z$-axis.
Therefore, the dynamics of $|\Delta_0|$ is dominated by the hole burning on $k_z$ axis, which explains the agreement between the oscillation frequency of the pairing amplitude $|\Delta_0|$ and the hole-burning frequency on the $k_z$-axis. 
The successively created hole-burning region on the $k_z$-axis has smaller extent in momentum space than the previous one due to the presence of hole-burning region created off the $k_z$-axis (see \reffig{polar_forward_2d}) and $\Delta_0$ will have a decaying amplitude in time.
A similar argument can be applied to the cases of backward quenches.


{\emph{Conclusion.--}} 
We have discovered the novel dynamics of a hole-burning and a particle-peak in the momentum occupation and the emergence of a vortex-ring structure in pair amplitudes. Our work also clarifies the mechanism of these dynamics and figure out the important role of quasi-bound (resonant) state in the BCS regime.
The hole-burning/particle-peak in momentum distribution has quite distinctive dynamical behavior which might be observed in the time-of-flight experiments \cite{Mukaiyama:private}.

We thank A. Bulgac, Q. Chen, B. Malomed, T. Mizushima, and T. Mukaiyama for helpful discussions.
This work was supported by the Max Planck Society, MEST of Korea, Gyeongsangbuk-Do, Pohang City, for the support of the JRG at APCTP, by Basic Science Research Program through NRF by MEST (Grant No. 2012R1A1A2008028), and by IBS through Project Code (IBS-R024-D1). GW is also supported by the Zhejiang University 100 Plan, by the Junior 1000 Talents Plan of China, and by NSF of China (Grant No. 11674283). Part of the computation was supported by the National Institute of Supercomputing and Network/Korea Institute of Science and Technology Information with supercomputing resources (KSC-2013-C3-004 and KSC-2015-C3-022).


\end{document}